\documentclass[prl,twocolumn,aps]{revtex4-2}

\usepackage{color}
\usepackage{graphicx}
\usepackage{subfigure}
\usepackage{afterpage}
\usepackage{dcolumn}% Align table columns on decimal point
\usepackage{bm}% bold math
\usepackage{color}
\usepackage{float}
\usepackage{amsmath}
\usepackage{bm}
\usepackage{amssymb}
\usepackage{siunitx}
\usepackage{cleveref}
\usepackage{natbib}
\usepackage{multibib}

\usepackage{natbib}

\def\url#1{\expandafter\string\csname #1\endcsname}

\newcounter{defcounter}
\usepackage{amsthm}

\begin{document}

\preprint{APS/PRL}

\title{Chaotic waves serve as universal pattern generators}% Force line breaks with \\

\author{Sergey A. Vakulenko,$^{1}$ Ivan Sudakow,$^{2,*}$ John Reinitz,$^{3}$ and Dmitry Grigoriev$^{4}$}
\affiliation{$^1$Institute of Problems in Mechanical Engineering, Russian Academy of Sciences, Bolshoj pr., 61, St.\,Petersburg, 199178, Russia\\
$^2$Department of Physics, University of Dayton, 300 College Park, Dayton OH, 45469, United States\\
$^3$ Departments of Statistics, Ecology and Evolution, Molecular Genetics and Cell Biology, University of Chicago, 5747 South Ellis Avenue, Chicago IL, 60637, United States\\ %reinitz@galton.uchicago.edu \\
$^4$ CNRS, Math\'ematiques, Universit\'e de Lille, Villeneuve d'Ascq, 59655, France}

%\email{Corresponding author: isudakov1@udayton.edu}

\begin{abstract}
Excitable media are prevalent models for describing physical, chemical, and biological systems which support wave propagation. In this letter, we show that 
%these media exhibit new wave propagation phenomena. 
the time evolution of the medium state at the wave fronts can be determined by complicated chaotic attractors. Wave front dynamics can be controlled by initial data choice. Building on this groundwork, we show that there is a mechano-chemical analog of the Universal Turing machine for morphogenesis problems. Namely, a fixed mechano-chemical system can produce any prescribed cell pattern depending on its input (initial data). 
This universal mechanism uses fundamental physical effects: spontaneous symmetry breaking with formation of many interfaces (kinks), which interact non-locally via a fast diffusing reagent.  This interaction creates chaos.
We present algorithms allowing us to obtain a prescribed target cell pattern.

%{In this paper, we consider  excitable media. We show that in these media the propagation of waves with chaotic and time periodic fronts is possible, as well as waves, which transfer associative memory. The time evolution of medium state at the wave fronts are determined by attractors. We can completely control these attractors by initial data choice. By those  results we show that there is a mechano-chemical analogue of Universal Turing machine. Namely, a fixed mechano-chemical system based on fundamental physical effects is capable to produce any prescribed patterns depending on its input (initial data).

\end{abstract}

\maketitle

%\tableofcontents

%\section \label{intro}

%Beginning with the seminal  Turing paper \citep{Turing}, many authors use
%reaction-diffusion models to describe pattern formation in
%biology, physics and chemistry.  Such models are capable 
%to describe excitable media, which exhibit many interesting phenomena.

\begin{figure}[!tbp]
%\vskip-0.5truecm
\centering
\includegraphics[width=0.3\textwidth]{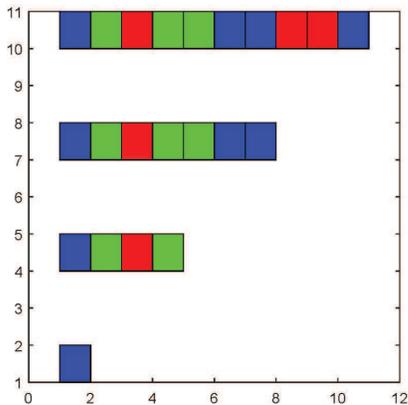}
\caption{\small{The generalized French Flag model in terms of wave morphogenesis. A cell pattern can be considered as a string in the alphabet (red, blue, or green).  Universal Turing machine can print any strings on a tape. Chaotic waves can do the same: they propagate along the $x$-axis and transform non-differentiated cells into differentiated ones.  We can obtain any prescribed string by a choice of initial data $z_0$ and the kink number $N$. 
} 
\label{fig2}}
 \end{figure}

{\it Introduction.} --- We propose a model of an excitable medium that can generate waves of a new kind.  These waves consist of interacting narrow fronts. The evolution of the coordinates that define the localization of those fronts, is governed by a dynamical system. The key point is that we can control the attractors of these dynamical systems by positional information stored in spatially distributed initial data and by the choice of a few of parameters. These attractors may be chaotic and of high dimension. We show that this effect has important biological consequences.  As an example, we consider applications to morphogenesis, in particular,  to cell differentiation problems. We resolve the cell pattern generation problem: imagine an arbitrary string of cells of different types located along the $x$-axis (this might model $1D$-organisms, like a worm, or a segmented embryo, see Fig.  \ref{fig2}). The cell pattern can be generated by our excitable medium, and we present an algorithm for how to do so.  

To better understand our approach to the cell differentiation problem in more detail, recall two fundamental biological concepts. An organism can be represented as a pattern consisting of different cells (see Fig.  \ref{fig2}).  The cells are “specialized”, i.e., each type of cell performs a unique and special function and each of the order of $100-200$ different types of cells in multicellular organisms has different structures, sizes, shapes, and functions. The famous Turing instability approach \citep{Turing} allows us to obtain periodical layered patterns, such as zebra stripes, however, we would like to explain more complicated observed structures. To this end, the concept of positional information was proposed by Wolpert \citep{Wolpert}. Both approaches, Turing's and Wolpert's, assume that morphogens, special reagents, can change cell states. 

%\vspace{-0.1em} 
Our new idea is that the waves with complex 
%dynamically 
evolving fronts can perform cell differentiation in a dynamical way. This allows
us to create any pattern not just periodic ones. The waves transfer a family of morphogenes, which change the cell states and produce cell differentiation. In contrast with Wolpert's gradient model, the wave act at long distances and 
can transfer dynamical information contained in an attractor. 

The main idea of the pattern generation mechanism is as follows. We restrict ourselves to one-dimensional layered patterns (a generalization to multidimensional cases will be presented in future papers). Consider the pattern shown on Figure \ref{fig2}. That pattern can be considered as a string of cell types (blue, green, red). Our aim is to create any such string. Note that a universal Turing machine (UTM) may print any string. A UTM includes a head and a tape, the states of the head form a finite set. The head moves along the tape and prints symbols.  Our medium generates waves, which move along the $x$-axis, and prints cells of different types. The type choice depends on the state of the wave front, defined  by a chaotic hyperbolic dynamics. Here we use the beautiful idea from C. Moore \cite{Moore1, Moore2} on simulation of TM's by chaotic dynamical systems. It is based on so-called Bernoulli shifts, chaotic dynamics can be encoded as a shift on a discrete set of symbols. So, the states of the waves can be encoded by a finite partition of all possible morphogen states. We present two variants of patterning algorithms, the first gives us a rigorous method to resolve any $1D$ problems of pattern generation, and the second is a simplified variant that works well in numerical simulations.

These results show that there are media that function as analogs of UTM's. A UTM can make all computations, which can be done by other TM's, and so, UTM's generate all possible string outputs when we vary their input. In our case, we have a fixed (up to a few parameters to adjust) spatially extended system, which, depending on initial data, generates all possible layered cell patterns.  Note that UTM's admit a short description \cite{Rog}.

So, our results show that simple mechanochemical systems can serve as  Universal Generators of spatio-temporal patterns  (UPG).  Thus they can be considered as analogs of UTMs. A UTM obtains a program as an input and performs computations prescribed by that program.  In our case, the input of our UPG is determined by spatially distributed initial data localized in a narrow domain.

Cell differentiation waves are proposed in \citep{Gordon1}, see also \citep{Gordon2}.  Cell killer waves are found in \citep{Apopt}. Apoptosis (programmed cell death) propagates through the cytoplasm as self-regenerating trigger waves, which spread without slowing down or petering out. Cell differentiation waves in Drosophila morphogenesis are found experimentally and investigated in \citep{NatWaves}, where, moreover,  a conceptual mathematical model is proposed, which involves reaction and diffusion, and exploits mechano-chemical effects, where chemical reaction terms are linear and quadratic.  The model  \citep{NatWaves} describes the time evolution of concentrations of free Fog ligand, bound-receptor Fog, and MyoII protein. 

Waves of cell differentiation are studied experimentally in \citep{PLOS2019}, where  it is indicated that signaling patterns may be dynamic, and cells may use various strategies to interpret these dynamics.  To investigate this dynamical mechanism, in \citep{PLOS2019} WNT and Nodal signaling pathways are studied. BMP signaling triggers waves of WNT and NODAL signaling activities, which move toward the colony center at a constant rate. It is shown that it is inconsistent with reaction-diffusion-based Turing models, suggesting that neither WNT nor NODAL forms a stable spatial gradient of signaling activity. So, the experiments and theoretical models show that, at least in certain situations, the morphogenesis proceeds with the help of waves, while the celebrated Turing instability does not work \cite{PLOS2019}.  However, the pathways involved in the wave dynamics are extremely intricate.

Similarly to \cite{PLOS2019, NatWaves}, in our model we use reaction and diffusion, and also linear elastic waves but we also implement into our model the scalar Ginzburg-Landau (GL) equation with a small gradient term. That equation describes bistability,  and spontaneous layered patterning. The GL equation simulates a trigger mechanism, which in real biological systems is generated by positive feedback loops in gene regulation networks  (those loops are detected in killer waves \citep{Apopt}). This extends possibilities in a formidable way:  spontaneous symmetry breaking creates complicated dynamical information and transfers that information through active media. 
 
 Let us outline our model. It consists of three equations. The first equation is a weakly perturbed Ginzburg-Landau (GL) equation for a scalar order parameter $u$. We suppose that the coefficient $\epsilon^2$ at the gradient term in the corresponding energy is small. It is well known that the non-perturbed GL equation has asymptotical solutions describing kink chains, where $i$-th kink is localized at $x=X_i(t)$.  Kinks are narrow topological defects (of width $O(\epsilon)$) with the charge $(-1)^i$  describing a symmetry breaking: a separation of the entire domain on subdomains along $x$-axis,  where $w \approx \pm 1$. Note that the direct interaction between kinks is exponentially small and therefore such a solution is correct within an exponentially long time $O(\exp(-c_1/\epsilon))$ while kinks are separated \cite{Carr}. Furthermore, we use a simple perturbation, which makes the kink chain move as a whole at a low constant speed $\kappa$. The following equation describes the reaction-diffusion dynamics of $v$-reagent, where the order parameter $u$ is involved. Reagent $v$ diffuses fast. 
The kinks interact with the fast reagent and the reagent $v$ acts on $u$, which that produces feedback and non-local non-direct kink interaction. 
We show that under an appropriate choice of system parameters the dynamics of the kink coordinates 
$X_i$ can be described by the Hopfield system with continuous-time and non-symmetric interactions. It is well known that such Hopfield systems exhibit a remarkable universality
property \cite{Vak2}: they can generate any structurally stable (hyperbolic) dynamics. Such dynamics may be chaotic
(the best known examples are given by Anosov flows and Smale horseshoes \cite{Ruelle1, Katok}). Following \cite{Moore1, Moore2} we can use this chaos to simulate Turing machines and we apply it to program pattern formation. 

We would like to note that curved chaotic fronts can also be described by the Kuramoto-Sivashinsky (KS) equation \cite{Siv, Kur}.  In our case,  a physical mechanism of the chaos generation is absolutely different: instead of curvature effects,
we use a non-local kink interaction via coupling with a fast diffusing reagent. While most of the known results for the KS model are numerical (see, for example,  \citep{Pathak}), our model is analytically tractable and there is an algorithm to control the wave front dynamics.

{\it The model and its properties.} --- %\label{Erd}
% \subsection{Model} 
 The model consists  of a reaction-diffusion part, a hyperbolic equation, and a scalar Ginzburg-Landau equation for an order parameter $w$:  
\begin{equation}
   u_t  =\frac{\epsilon^{2}}{2} \Delta u + u- u^3 -\kappa u_x + \gamma v,  \label{GLE}
\end{equation}
\begin{equation}
  { v}_t    = \Delta { v}  + z u_x,
\label{OB2M}
\end{equation}
\begin{equation}
   z_t + \kappa z_x=0. \label{Hyp}
\end{equation}
Here $\gamma, \kappa>0$ and $\epsilon>0$ are small parameters, $u=u(x,y,t)$ and ${v}(x,y,t)$ are   unknown functions  defined on $\Omega \times \{ t \ge 0 \}$,
 $\Omega$ is the strip
$(-\infty, \infty) \times [0,1] \subset {\bf R^2}$.  Eq. (\ref{Hyp}) for $z$ can describe elastic (mechanical) effects, and 
the deformation $z$ affects $v$ via a quadratic nonlinearity.
To simplify the problem,
and  bearing in mind further the propagation of waves,
we set the periodic boundary conditions 
\begin{equation}
  v(x, y, t)={ v}(x+ 2\pi, y, t), 
\quad { u}(x, y, t)={ u}(x+ 2\pi, y, t). \label{period}
\end{equation}

At the boundaries $y=0$ and $y=1$ we set the zero Dirichlet  conditions
for  ${v}$:  
\begin{equation}
 { v} (x,h, t) = { v}(x, 0, t)=0
\label{Dir}
\end{equation}
and  the zero Neumann condition for $u$
\begin{equation}
  u_y(x, y, t)\Big \vert_{y=0, 1} =0.
\label{Neum}
\end{equation}
The initial conditions are given by smooth functions $u_0, v_0$ and $z_0$, for example, 
\begin{equation} \label{initdata}
   z(x,y,0)=z_0(x,y),
\end{equation}
and similarly for $u, v$. The function $z_0$ plays a key role in long time behaviour control.

The key difference between this system and the model of \cite{NatWaves} is the presence of the GL equation (\ref{GLE}), which describes phase transitions and layered patterning.  So, we can take into account basic mechanical, chemical, and physical effects, and we think that this model is the most efficient among all those providing the effects described in the manuscript.
Note that our model is two-dimensional that is important for the control of large time dynamics. To obtain analogous results
in one-dimensional case, we have to use a number of reagents replacing a single eq. (\ref{GLE}) by 
a reaction-diffusion system.

{\it Asymptotic solutions and mechanism of chaos onset.}--- %\label{asym}
In this model, chaos appears as  a result of a non-local kink interaction. 
For each integer $N$ and sufficiently small $\epsilon, \kappa>0$  and $\gamma$  there exist  solutions
describing interaction of $N$ kinks.  The $u$-component
   of these solutions are perturbed $2\pi$-periodic in $x$ kink chains 
   $U_N(x, X(t))$ consisting of $N$  kinks well localized at points  $X_i(t) - \kappa t$,
   where $ X_1 > X_2 < ... > X_N> \delta_0 >> \epsilon$ are slowly evolving in time relative kink  coordinates. 
   Analogous kink solutions for (\ref{GLE}) are described first in \citep{Carr}.
Such solutions are metastable and exist while kinks are well separated, 
and  the kink existence time interval $I_{\epsilon}$ is of the order  $\exp(-c_0 \epsilon^{-1})$ \citep{Carr}. 
So, our  solutions have the form 
\begin{equation} \label{asym21}
    u(x,y,t)=U_N(x,t) 
    +\tilde u(x, y, X(t)),
\end{equation}
\begin{equation} \label{asym22}
    v(x,y,t)=V_N(x, y, t) 
    +\tilde v(x, y, t), 
\end{equation}
where $\tilde u, \tilde v$ are small corrections with respect to the main terms $U_N$ and $V_N$. The following relation
is important:
%\begin{equation} \label{WN}
%   \frac{\partial {U_N}(x, t)}{\partial x}=\epsilon^{-1} \sum_{i=1}^N (-1)^i \cosh^{-2}(\epsilon^{-1}(x - X_i(t)-\kappa t)),
%\end{equation}
  \begin{equation} \label{WNV}
V_N(x, y,  t) =\sum_{i=1}^N X_i(t) W_i(x-\kappa t, y), 
\end{equation}
where $W_j$  are smooth functions. The function $W_i$ defines a response of $v$-reagent  to the excitement generated by $i$-th kink. In turn, the $v$-reagent acts on kinks via the small perturbation $\gamma v$ in Eq.
(\ref{GLE}). So, we obtain a feedback and a non-local nonlinear interaction between the kinks, which is much stronger than  exponentially small interactions between nearby kinks.
For an appropriate choice of the small parameters $\epsilon, \gamma, \kappa$, and the initial data $z_0(x,y)$ one can show that, up to small corrections, the time evolution of kink coordinates $X_i$  is governed by the  time continuous Hopfield system 
\begin{equation} \label{maindyneq}
    \frac{dX_i}{dt} = \sum_{j=1}^N  K_{ij} \sigma(X_j-h_j) - \lambda X_i, 
\end{equation}
where $\sigma$ is a smooth sigmoidal function, the matrix
${\bf K}$ with entries $K_{ij}$ defines an interaction between $X$, $h_j$ are thresholds and $\lambda>0$. The form of this system
depends on parameters ${\bf P}$,  ${\bf P}=\{ {\bf K},  N, h,\lambda\}$.  The matrix ${\bf K}$
 and $\lambda$ are linear  functionals of initial data
 $z_0$. The key point is that by variation of $z_0$
 we can obtain any given ${\bf K}$ (not necessarily symmetric, see SM).
 
The Hopfield systems with general non-symmetric interactions $K_{ij}$  enjoy remarkable properties. We know that multilayered perceptions can approximate any output (Theorem on Universal Approximation). By that basic result, one can show
that the Hopfield system  has the property of Universal dynamical approximation. Namely, they can simulate, within any prescribed accuracy, any finite-dimensional dynamical systems (see \citep{Vak2, switch} and SM). This simulation works via hidden slow variables, which appear in the Hopfield dynamics under an appropriate choice of ${\bf K}$. Then, that matrix defines an interaction between slow and fast variables. As is typical, in such slow-fast systems, the slow variable dynamics captures the entire system's long-time behavior. By parameter ${\bf P}$ we can completely control the slow dynamics (up to small smooth corrections).   

For example, suppose we would like to simulate the Lorenz dynamics within accuracy $\delta$. Then we can adjust parameter ${\bf P}$ in such a way that (\ref{maindyneq}) becomes a slow-fast system, and the slow part dynamics is defined by the $\delta$-perturbed Lorenz system.
This simulation holds, in general, on large time intervals, but if the attractor of the prescribed system is structurally stable (for example, hyperbolic), i.e., does not change its topology under sufficiently small and smooth perturbations, then for small $\delta$-the simulating Hopfield dynamics is the same (up to topological equivalency of trajectories). Roughly speaking this means that system (\ref{maindyneq}) can simulate all hyperbolic dynamics, for more precise formulation see \citep{Vak2} and SM). These facts lead to the results described in the coming section. Note that a connection between the neural network Hopfield model and reaction-diffusion systems was first discovered in \citep{Edw}, see \citep{Vak2} for a rigorous proof.

{\it Formation of cell patterns.} --- To describe patterns consisting of differentiated cells and cell differentiation via the reagent $u$, we use the model, which follows the biological ideas \citep{Turing, Wolpert} outlined in the introduction. Consider, for simplicity,  two cell types, say,
red and blue cells. We encode them by $1$ and $2$, respectively (the generalization for a larger number of cell types is quite straightforward).  We assume that cells occupy strips of the same small length $\delta_c$ forming a layered pattern along the $x$-axis.   We thus have $M=[L/\delta_c]$ equidistant layers. The output cell pattern can be considered as  a binary string $s_{out}$: $s_{out}=\{a_1, a_2, ..., a_M \}$, where $a_i$ is either $1$, or $2$. We also introduce a state $0$.  The state zero corresponds to cells that are not yet differentiated.  
 
Next, we describe how the cell pattern can be produced in our model. 
The cell pattern is a result of terminal differentiation which goes by morphogens. Suppose that the $u$ is a  morphogen.  It is natural to assume that cells interpret morphogen signals by averaging in space and time. For simplicity, we assume that this interpretation goes through linear convolution operators, which act on the $u$-pattern (see SM). So, the cell obtains information about kink coordinates $X$  at the moments when the kinks reach the cell. Let us consider how this information can be used. The range of all possible values $X$ will be denoted by $\Pi$.  We introduce the partition of $\Pi$ consisting of disjoint subsets  $\Pi_k$, $k=0,1, 2$ such that their union is $\Pi$ and each subset has an open interior. This partition has a simple meaning: we encode the continual space of wave states by a discrete code.  The set  $\Pi_0$ corresponds to non-differentiated cells, the set $\Pi_k$ with $k >0$ corresponds to cells of $k$-th type.  We encode kink states $X=(X_1, ..., X_N)$ by  functions $Z(X)$. The coding function $Z(X)$ takes the value $k$ if $X \in \Pi_k$.  Let $u(x, y, t)$ be the asymptotic kink solution. Then the output string $s_{out}[u]$ can be  defined as follows: $j$-th element of the string is $k$, if $Z(X(t_j))=k$, where $t_j$ is the moment when kink chain wave reaches $j$-th cell.  This construction replaces thresholds in the Wolpert positional information approach, but in our case, this information is transferred in the cells by waves instead of gradients. We refer to $s_{out}[u]$ as wave cell differentiation operator, for more details see SM.  

Note that at the moment $t=t_j$ the corresponding cell accepts $k$-th state and does not change its type anymore.  We assume here the  biological fact that typical cells do not change their cell types after terminal differentiation when they acquire their specialized type. At $t=0$ all cells are in an indefinite state $0$.  
 %We discuss the robustness of this construction in sect. \ref{}.
 %our model has the property of dynamic approximation: when we change its parameters, it simulates any dynamics
 
{\it Main results.}---Concluding the ideas presented above we formulate the following statements. 

{\it On dynamical complexity: Kink dynamics of our model has the property of universal dynamical approximation.}

This means that when we vary the model parameters, initial data  and the kink number,  kink coordinate dynamics can generate all possible kinds of structurally stable large time behavior (up to topological equivalency). Since hyperbolic dynamics is persistent \citep{Katok}, kink motions generate all hyperbolic dynamics. Hyperbolic dynamics may be chaotic \citep{Katok, Ruelle1}, and further, we show how hyperbolic chaos generates all possible $1D$ layered patterns. It can be done  by an algorithm, which allows us to obtain a prescribed cell pattern.
%\begin{figure}[!tbp]
%\vskip-0.5truecm
%\centering
%We can extend   the model  (\ref{cellnum}) assuming that the output $F_{cell}$ is determined by multilayered perceptrons or even  deep networks
%(such models can be used to describe enhancers, see, for example, \citep{enhancer, Reinitz2020}).  

The next statement unwraps the main problem of cell pattern formation.

{\it On the cell pattern generation problem: Let $s=\{a_1, ..., a_M \}$ be a prescribed string  of cell types, $a_i \in \{1, 2\}$. To  find parameters $\epsilon, \kappa, \gamma$ and initial data $u_0, v_0, z_0$ such that the corresponding solution of Initial Boundary Value Problem (IBVP) defined by eqs. (\ref{GLE})-(\ref{Hyp})  and conditions (\ref{Neum})-(\ref{initdata})
 satisfies
\begin{equation} \label {cellform}
    s_{out}[u]=s,
\end{equation}
where $s_{out}[u]$ is the wave cell differentiation operator.}

The last statement can be formulated as follows. 

{\it The  pattern generation problem has a solution}. 

We describe algorithms to resolve this problem in the coming section.
%However, the set of initial data $z_0(s)$ generating a given $s$ becomes more and more narrow for long and complex strings, so, the %search of solution is a hard problem. 

{\it Pattern generation.}--- We propose algorithms to solve the pattern generation problems based on celebrated results of dynamical system theory on hyperbolic sets, in particular, the existence of Markov partitions that implies the correspondence between maps on invariant hyperbolic sets and Bernoulli shifts \citep{Katok, Moore1, Moore2}. The idea of the algorithm can be outlined as follows. We first encode a cell pattern as a string in an alphabet ${\mathcal K}_c$ of cell types. The algorithm input is  then a string $s_{out}=\{a_1 a_2 ... a_M\}$ of symbols from ${\mathcal K}_c$ (see Fig. \ref{fig2}). We would like to produce such a string. We know that a UTM can print that string: the UTM head moves along the tape and prints.  Similarly, our wave moves along the $x$-axis and prints different cells. Although states $X$ of that wave lie in a bounded domain of ${\mathbb R}^N$, we can make a partition of that domain
to encode the wave states. Then the wave becomes an analog of the UTM head. Here we use Bernoulli shifts and the same idea that allows realizing TM's by dynamical systems, see SM for more details.
 
Note that the algorithm is based on the well-known biological fact that cells (except for stem cells) are not capable of further differentiation.  When a wave comes to an area occupied by a cell, it changes its type (depending on the amplitude of the wave), that is, it makes differentiation and after that, the cell no longer changes.

So, we conclude  that there is a universal reaction-diffusion system, which can produce any cell phenotypes depending on initial data and a few parameters, i.e., we can obtain a needed
final (terminal) phenotype.  

The chaotic hyperbolic attractor can be taken, in principle, in an arbitrary way, however, it is natural to take a low dimensional one. 
The choice of the Markov partition depends on the coding scheme, which we use for cell types. 

Our chaotic attractors generate strings within time intervals, but by waves described above, we can obtain a generation along the $x$-axis. 

%\subsection{Simplified algorithm and numerical simulations}%\label{numres}

{\it Numerical example and simplified algorithm.}--- As an example of the algorithm application, let us consider how to create the pattern in Fig. \ref{fig2}. Numerical simulations show that the sophisticated algorithm stated above can be strongly simplified. We can, instead of the Markov partitions, use almost arbitrary partitions of phase space on disjoint subsets. We consider first how to generate layered pattern like the famous French flag  by waves instead of gradients. Consider the pattern in Fig. \ref{fig2}  consisting of $4$ layers: blue, green, red, and again green.  Suppose for simplicity that all layers of the cell pattern have the same width then the string corresponding to that pattern is $s=(1, 2,  3, 2)$.
We take the Lorenz system for variables $q=(q_1, q_2, q_3)$ with the standard choice of parameters to produce a chaotic attractor $\Gamma$. Further, we find the Hopfield system such that the kink coordinates $X$ evolve according to weakly perturbed Lorenz system, 
$X(t) \approx X(q(t))$.

Let us introduce $q_{min}=\min_{q \in \Gamma} q_1$,
$q_{max}=\max_{q\in \Gamma} X_1$, and $\Delta q=q_{max}-q_{min}$.
Then we take the partition
$E_1=[q_{min}, q_{min} +\Delta q]$, $E_2=[q_{min}+\Delta q, q_{min} +2\Delta q]$ and $E_3=[q_{min}+2\Delta q, q_{max}$.
Then one can check numerically that there exist points
$q(0)$
on the Lorenz attractor and $\Delta T$ such that $X(\Delta T)
\in E_1$, $q(2\Delta T)  \in E_2$,   $q(3\Delta T)  \in E_3$ and $q(4\Delta T)  \in E_2$. Here $E_1, E_2, E_3$ correspond to
blue, white and red cells, respectively. The partition of $X$-space, which define cell differentiation,
is formed by ranges of $E_k$ under the map $q\to X(q)$.
So, we obtain the layered aperiodic $1D$-pattern consisting of four layers (see Fig. \ref{fig2}, the second row from bottom). The same construction allows us to obtain more complicated patterns, for example, consisting of five and more layers.  Note one can take other sets $E_k$ so the choice of the partition is almost arbitrary. However, the longer the cell pattern becomes, the smaller the set  of starting points $q(0)$ will be, and thus it is more difficult to find that set. 
 
This simplified variant of the algorithm can be analytically explained under the assumption that the dynamics on the attractor is strongly mixing (see SM, subsect. \ref{simpalg}).  Moreover, this variant is robust with respect to the choice of partitions. However, the sophisticated algorithm  with Poincar\'e map has an advantage: by the Bernoulli shifts and the Markov partitions, we can find the set of initial data $q(0)$ and the corresponding initial kink coordinates $X(q(0))$ in an explicit way. The pattern  generation by the simplified algorithm can be observed in a video, see  \citep{vid}.

\begin{figure}[!tbp]
%\vskip-0.5truecm
%\centering
\includegraphics[width=0.3\textwidth]{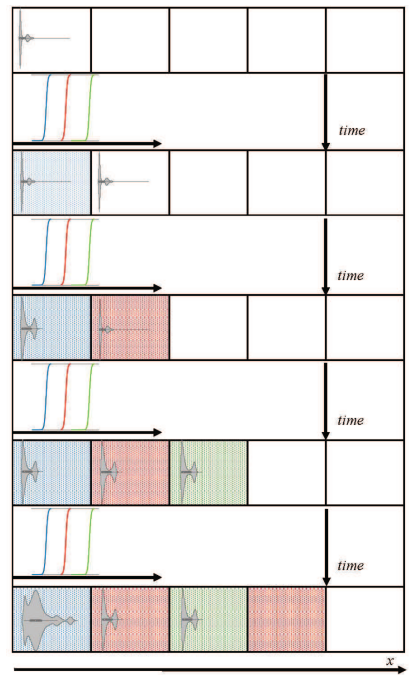}
\caption{\small{The active medium described in the paper can generate waves, which can transfer a complicated time behaviour. A cell colony that has created such a medium (or is immersed in it) can have important selective advantage.
For example, suppose that a cell, a member of that colony, finds a complex dynamical adaptive answer to
a ecological challenge (the top row, the first cell). Then this answer can be transferred to other cells by 
the waves, and thus the whole colony obtains an ability to survive. Moreover, it is shown that the wave front dynamics is defined by the Hopfield networks, so, those waves also may transport
associative memory.
} 
\label{fig3c}}
 \end{figure}

{\it Conclusions.} --- 
A key component for achieving functionally stable multicellular
structures is a physical embodiment. Any relevant model of the evolution of multicellular organisms should actually take into account  basic physical mechanisms.
It is shown that there exists a simple physical model defined by three equations with quadratic and cubic nonlinearities which create the chaotic waves of a new type. These waves have fronts, which can be interpreted as "moving" attractors and they can transfer information since dynamical systems with a complicated behavior can simulate all Turing machines
\citep{Moore1, Moore2}.  
Such waves can transfer information in space, for example,  innovations. Following \citep{RevMorph} one can say that excitable media can create programmed and self-organized flows of information.  Propagation of complicated information, which may seem to be the exclusive prerogative of human society is possible in simple physical media.
These results can be applied, in particular, to cell differentiation problems. New pattern formation
mechanism described here can produce any target cell $1D$ patterns. 

Physical processes mobilized by genes can establish morphological templates. Most animal body plans and morphological motifs arose between 500 and 700 million years ago, during relatively brief periods of innovation. The genes,  whose products control morphogenesis and pattern formation, were present in the unicellular ancestors of the animals;  billion years of evolution failed to generate substantial additional morphological novelty. The work \citep{NewmanS} reconciles these facts by proposing that chemically and mechanically active  media can create the main motifs of animal forms. 
Our results support this concept of physical determinism in development.  We also think that the proposed pattern generation mechanism can appear in other applications, for example, in ecology and economics.

The active media that generate complex  waves are simple and the generation mechanism involves fundamental physical and chemical effects of physics. A few genes is sufficient to correctly encode that mechanism. Therefore, it is natural to expect that such media could appear as a result of biological evolution.
One can imagine, for example, such a model (see Fig. \ref{fig3c}). Consider a cell colony that must adapt to a new environment and develops products necessary for survival. It is clear that a colony, where it is possible to transfer complex adaptive innovations from one cell to another, has a clear selective advantage. This transmission can be done by means of the waves, studied in this paper, and these waves can not only transmit simple information, but they can also transfer complex behavior (which can be described by an attractor or a Turing machine, or a neural network with associative memory), similarly to human society.

%\section*{Acknowledgments} 

 %First of all, we are grateful to Referees for useful and interesting remarks.
 {\it Acknowledgments.}--- S. V. and J. R. are supported by the grant of National Institutes of Health (NIH) 2R01 OD010936. IS gratefully acknowledges support from the Division Of Physics at the  U.S.  National  Science  Foundation  (NSF)  through  Grant PHY-2102906. 
\bibliographystyle{apsrev4-2}
\bibliography{referenceT}

%\setcounter{page}{1}
%\newpage
%\vspace{5cm}
%\cleardoublepage

%\appendix{EvksatF}

%\section{Supporting Information}
\clearpage
\section{Supplementary Material} \label{SI}

\section{Asymptotic solutions to system
(\ref{GLE})-(\ref{Hyp}) describing complex waves} \label{waves}

\subsection{Kink chains}

Let us describe kink chains following \citep{Carr}. Let first $\kappa=0$.
Let $X_j$, $j=1, \ldots N$  be coordinates of the kinks in the interval $(0, 2\pi)$. We suppose 
that $0 < X_1 < X_2 < ... < X_N <2\pi$. Let $dist_{+}(X)=\max_i {X_{i+1}- X_i}$ and
$dist(X)=\min_i {X_{i+1}- X_i}$. We assume 
\begin{equation} \label{kinkdist}
\epsilon << \delta_1 < dist(X)  \le  dist_{+}(X)   << \delta_c,
\end{equation}
i.e., maximal and minimal distance between kinks are small. The minimal distance is much more than the characteristic diffusion length
$\epsilon$ and the maximal one is much less than the cell size $\delta_c$. We need these assumptions to have an asymptotics solution in the form of a kink chain and also to construct a simple cell pattern generation operator (see subsect. \ref{TuringP}). 

The kink chains can be obtained by 
 $2\pi$-periodical in $x$ functions $\bar U_N(x, X)$, which have the following form. Inside narrow intervals
$I_{i, \epsilon}=\big(X_i - \epsilon^{1/2}, X_i + \epsilon^{1/2}\big) $,
$\bar U_N(x, X)$ has the form $s_i \tanh(\epsilon^{-1}(x-X_i)$, where $s_i=(-1)^i$ are topological charges and
$i=1, \dots, N$. Outside of intervals
$I_{i, \epsilon}$ the function $\bar U_N(x, X)$ is exponentially close to $\pm 1$ and this function is a smooth function of $x$. 
For $\gamma =0, \kappa =0$ we have a set of solutions $u$ of (\ref{GLE}), which have the form  $u=\bar U_N(x, X(t))+ \tilde u$,
where $\tilde u$ is a small correction and the kink coordinates $X_j(t)$ evolve in time exponentially slowly:
$|dX_i/dt|= O(\exp(-c_0 \epsilon^{-1} dist(X))$ 
\citep{Carr}. These solutions are correct while $dist(X) >> \epsilon$. Time evolution of $X$ is a result of exponentially weak
direct kink interaction. 
For $\kappa \ne 0 $ we obtain the  kink chain 
 travelling with a constant speed $\kappa$:
$
 U_N=\bar U_{N}(x-\kappa t, X)
$.

In the coming subsection we consider the case $\gamma >0$, where we have more complicated equations for $X$. In this case
there is a non-direct and non-local  kink interaction via coupling with $v$-reagent.

\subsection{Equations for kink coordinates} \label{eqsforX}

Solution of eq. (\ref{Hyp}) has the form
$z(x,y, t)=z_0(x -\kappa t, y)$. We substitute  $z$ into eqs.(\ref{GLE}), (\ref{OB2M})
and make variable change  $\tilde x=x -\kappa t$ 
(further we omit tilde in notation). Then we obtain the
following system 
\begin{equation}
   u_t  =\frac{\epsilon^2}{2} \Delta u   +  u - u^3  +
   \gamma v,  \label{OB1W}
\end{equation}
\begin{equation}
  { v}_t    = \Delta { v}  + z_0 u_x + \kappa v_x.
\label{OB2W}
\end{equation}

Further we find asymptotic solutions of that system under the following assumptions to 
 the small parameters: 

{\bf Assumptions to $\gamma, \epsilon$ and $\kappa$}

{\em Let $\epsilon>0$ be small enough and
\begin{equation} \label{asseps}
    \gamma < \epsilon^4, \quad  c_0 \exp(- c_1 \epsilon^{-1/2}) << \kappa << \gamma, 
\end{equation}
where $c_0, c_1$ are uniform in $\epsilon>0$.}

The main idea in choice of $\gamma$ is to conserve the planar structure of the kink fronts ( otherwise it is impossible 
to describe kink chains by coordinates $X_i$, and it is necessary to take  into account the front curvature). The  parameter
$\kappa >0$ should be small as well in order to obtain quasistationary solutions of  eq. (\ref{OB2W}).
The condition $c_0 \exp(- c_1 \epsilon^{-1/2})<< \gamma$ is neccesary to ensure domination  of non-local
kink interaction via $v$-reagent  with respect to direct kink one.

 The subsequent statement follows works \citep{Carr, Vak2,  VakVol, Vak16} with small modifications. Our first goal is to derive equations for $X_i(t)$. 
Eqs.  for $X_i(t)$   can be derived by  a standard perturbation approach for small $\gamma>0$
(see, for example, \citep{Carr, VakVol}).   For sufficiently small $\gamma>0$ one has
 \begin{equation}  \label{pertwave}
u(x, t)=U_N(x, X(t)) + \tilde u(x, t, \gamma),
\end{equation}
where $U_{N}(x, X)$ is the  kink chain (see above) and $\tilde u$ is a correction.
Then  the time evolution of $X_i$ is governed by the system
 \begin{equation}  \label{qevol}
\frac{dX_i}{dt}= \gamma \big(G_i(X, \gamma, \epsilon) + \tilde G_i \big),
\end{equation}
where
\begin{equation}  \label{qevol2}
\bar G_i=  -\frac{3}{2}\int_{0}^{2\pi}\int_0^1  v(x, y)  \cosh^{-2} (\epsilon^{-1} (x- X_i)) dxdy
\end{equation}
and  small corrections $\tilde G_i$ are uniformly bounded
$$
| \tilde G_i( X, \gamma, \epsilon)| < c \big(\gamma^s + \exp( \epsilon^{-1} dist(X)) \big),
 $$
where $s \in (0,1 ]$, $c, s$ are uniform in $\gamma$ as $\gamma \to 0$.

To explain  equations (\ref{qevol}) and (\ref{qevol2}),
let us remind the construction from \citep{VakVol}, which is the well known Lyapunov-Schmidt factorization.  To obtain the  dynamical equations for $X_i$, we impose the condition
\begin{equation} \label{orto}
\int_{0}^{2\pi} \int_0^1 \tilde u_i(x, y, t) \cosh^{-2} (\epsilon^{-1}(x- X_i(t)))  dxdy = 0
\end{equation}
for each $t$. These equations define $X$ uniquely for small $\gamma$ and bounded $v$. For the correction $\tilde u$ we obtain
\begin{equation} \label{veq}
\tilde u_t= {\bf  L}_{\epsilon}( X) \tilde u   +  H(\tilde u, X, \gamma),
\end{equation}
where ${\bf L}_{\epsilon} $ is the linear operator defined by
\begin{equation} \label{operator}
    {\bf L}_{\epsilon}  \psi=\frac{\epsilon^2}{2} \Delta \psi  + \big(1- 3U_{N}(x, X)^2 \big) \psi,
\end{equation}
and
$$
H(\tilde u, X, \gamma)= -\tilde u^3 - 3 U_N \tilde u^2.
$$
The spectrum of the operator ${\bf L}_{\epsilon}$ is well studied \citep{Carr}. 
Note that  ${\bf L}_{\epsilon}$  is a self-adjoint operator of Schr\"odinger type, which
 has a kernel consisting of $N$ eigenfunctions associated with kink Goldstone modes, which describe kink shifts.  
 
  Let $\Psi_j=\bar c_j \epsilon^{-1/2}\cosh^{-2} (\epsilon^{-1}(x- X_i))$,
  where the normalizing constants $\bar c_j$ provide $||\Psi_j||_{L_2(\Omega)}=1$. Note that
  $\bar c_j$ equal $\sqrt{3}/2$ up to exponentially small corrections. 
  
 We use the standard  notation
  $$
  \langle  f , g \rangle =\int_0^{2\pi} \int_0^1 f g dxdy, \quad || f||^2=\langle f, f \rangle.
  $$
   We need the following Lemma (see \citep{Vak16}).
  \\
  {\bf Lemma 1}{\em   For $X$ such that $dist(X) > \delta_1>0$
 one has 
  \begin{equation} \label{exp1}
     || {\bf L}_{\epsilon}  \Psi_j || < c_0 \exp(-c_1 \epsilon^{-1})
  \end{equation}
  and if $\langle \psi, \Psi_j \rangle =0$ for all $j$ then
  \begin{equation} \label{exp2}
     || {\bf L}_{\epsilon} \psi || \le - C_0 \epsilon^2 || \psi||,
  \end{equation}
  where all constants are uniform in $\epsilon$ as $\epsilon \to 0$.
  }
  \vspace{0.2cm}
  
 For a proof see \citep{Vak16}.

 So, the spectrum of ${\bf L}_{\epsilon}$ consists of $N$ exponentially small eigenvalues and 
  all the remaining spectrum of ${\bf L}_{\epsilon}$
lies in the  interval $(- \infty, -\delta_0 \epsilon^2)$, where $\delta_0>0$ does not depend on $\epsilon>0$.  This property implies the stability of the kink solutions on exponentially large intervals $I_{\epsilon}$ and allow us to
solve eq. (\ref{veq}). It can be done  by the standard perturbative methods because these equations involve  weak nonlinearities while the linear part is stable due to condition
(\ref{orto}), which  also implies dynamical equations (\ref{qevol}) for $X$.

Using Lemma 1, and eq. (\ref{veq}) one can obtain the following estimates for $\tilde u$ :
\begin{equation} 
|| \tilde u||  \le C_1 \gamma \epsilon^{-2} || v||, \quad \sup |\tilde u| \le C_2 \gamma \epsilon^{-2}\sup |v|.
\end{equation}
The constants in these estimates are uniform in $t$ and $\gamma, \epsilon >0$. More general estimates can be obtained by semigroup theory, see subsubsect. \ref{eqtueqtv}.

\subsection{Quasistationary solutions of (\ref{OB2M})}

Let us turn  now to equation (\ref{OB2M}) for $v$. 
This equation is linear with respect to both $v$
and $u$, $u$ is a sum $U_N + \tilde u$, where $\tilde u$ is small. The function $U_N$  depends on time via  slow variable $X(t)$
and does not depend on $t$ explicitly. Therefore, we can solve that equation  by a simple idea:
 we can freeze $X$ in eq. (\ref{OB2M}) assuming  that $X$ is just a parameter. Thus the main contribution to $v$ is given  by the function $V_N$ satisfying the equation
$$
\Delta V_N = \big(\kappa{V_N} + z_0 {U_N} \big)_x.
$$
We obtain 
\begin{equation}\label{VN}
    V_N=\sum_{j=1}^N W_j, 
\end{equation}
\begin{equation}\label{Wj}
    \Delta W_j - \kappa {W_j}_x= z_0(x,y) \epsilon^{-1} \cosh^{-2}(\epsilon^{-1}(x- X_j)).  
\end{equation}
For small $\epsilon>0$ the function $\epsilon^{-1} \cosh^{-2}((x- X_j)/\epsilon)$ is a good approximation of $\delta$-function
(up to a constant uniform in $\epsilon>0$). Moreover, it is clear then that in eqs. (\ref{Wj}) 
$z_0(x,y)$ can be replaced by $z_0(X_j, y)$. Let us denote by $\Gamma_m(x-x_0)$ the Green function of the one-dimensional boundary value problem satisfying the equation
$$
\frac{d^2 \Gamma_m }{dx^2} - m^2  \Gamma =\delta(x-x_0)
$$
and the $2\pi$ -periodical boundary conditions in $x$. 
%Note that for large $m$ we have asymptotics
%
Then we resolve 
(\ref{Wj}) by the Fourier method that gives
\begin{equation}\label{WjF}
     W_j(x, y, X_j) = \sum_{m=1}^{\infty} b_m \hat z_{m}(X_j) \sin(\pi m y)   \Gamma_m(X_j -x),
\end{equation}
where $
b_m= (1- (-1)^m)/(\pi m),
$
and $\hat z_{0, m}(X_j)$ are the Fourier coefficients of 
$z_0(X_j, y)$:
$$
\hat z_{m}(X_j) = 1/2 \int_0^1 z_0(X_j, y) \sin(\pi m y) dy,  
$$
where $m$ are positive integers.

\subsection{Hopfield system} \label{Hop}

Using relations (\ref{VN}),   eqs.  (\ref{qevol}), (\ref{qevol2}), and removing small  terms, we obtain
evolution equations for kink coordinates :
\begin{equation}\label{preHop}
     \frac{dX_i}{dt} =  \gamma G_i(X), 
\end{equation}
where
\begin{equation}\label{preHop2}
     G_i(X)=-\frac{2}{3} \sum_{j=1}^N  \int_0^1  W_j(X_i, y, X_j)dy.
\end{equation}
Our goal is to reduce this system to the Hopfield one. It can be done by a special choice of
$z_0(x,y)$, or, that is equivalent, of $\hat z_{m}(X_j)$.  First we substitute   formula (\ref{WjF}) into
(\ref{preHop2}). Then we have
\begin{equation}\label{preHop3}
     G_i(X)=-\sum_{j=1}^N   \sum_{m=0}^{\infty} \frac{  \hat z_{2m+1}(X_j) P_{ij,m}(X) }{3\pi(2m+1)^2},
\end{equation}
where
$$
P_{ij, m}(X)=\Gamma_m( X_j -  X_i).
$$
The main idea to simplify the  formula (\ref{preHop3}) 
for $G_i$ is as follows. Suppose that the
kinks oscillate at certain fixed points $\bar X_j$, i.e.,
\begin{equation}\label{Osc}
     X_i(t)=\bar X_i +  \tilde X_i(t),
\end{equation}
where $\tilde X_i$ are new unknowns. Suppose temporarily that $|\tilde X|=O(1)$ as $\rho \to 0$, where $\rho > 0$  is a small parameter.
(this assumption will be justified later).  We can achieve such behaviour of solutions under
a special choice of $\hat z_{2m+1}(X_j)$. Positions of points $\bar X_j$ may be arbitrary but the condition 
$$
\epsilon << \delta_1 < dist(X)  \le  dist_{+}(X)   << \delta_c,
$$
must be satisfied.

Namely, we
set
\begin{equation}\label{Osc1}
     \hat z_{2m+1}(X_j)= \xi_{mj} (\tilde X_j) +
     \eta_{mj} (\tilde X_j),
\end{equation}
\begin{equation}\label{Osc2}
     \xi_{jm} (\tilde X_j)= \rho M_{jm} \sigma(\rho^{-1} \tilde X_j -h_j),
\end{equation}
\begin{equation}\label{Osc3}
     \eta_{jm} (\tilde X_j)=S_{jm} \tilde X_j,
\end{equation}
where $\sigma$ is a smooth sigmoidal function, for example, 
$$
\sigma(z)=(1 + \exp(-z))^{-1}, 
$$
and where $M_{jm}, S_{jm}$ are unknown coefficients, which
must be matched appropriately.

We obtain then
\begin{equation}\label{preHop4}
     G_i(\tilde X)= \sum_{j=1}^N   \sum_{m=0}^{\infty}  
     \big(M_{jm} \sigma(\rho^{-1} X_j -h_j)+ S_{jm} \tilde X_j \big) P_{ij, m}(X).
\end{equation}

 Then we can simplify (\ref{preHop4})  that gives (up to terms of the order $O(\rho)$)
\begin{equation}\label{preHop5}
     G_i(\tilde X)= \sum_{j=1}^N   \sum_{m=0}^{\infty}  
     \big(M_{jm} \sigma(\rho^{-1} \tilde X_j- h_j)+ S_{jm} \tilde X_j \big)P_{ij, m}(\bar X).
\end{equation}

Further we use the following lemma. 
\\

{\bf Lemma II.}
{\em For each  $N\times N$ matrix ${\bf K}$ with entries $K_{ij}$
 there exist a number $M \ge N$ and coefficients $b_{jm}$  such that  
\begin{equation} \label{appr2}
       \sum_{m=0}^{M}   b_{jm} \Gamma_m(\bar X_j - \bar X_i) = K_{ij} \quad \forall i, j.
\end{equation}
}

{\bf Proof}.
 For unknown $b_{jm}$ we have a system of linear algebraic
 equations. For large $m$ we have asymtotics
 \begin{equation} \label{Green}
  \Gamma_m(x-x_0) =(2m)^{-1} \big(\exp(-m|x -x_0|) (1 + o(1)) \big), 
\end{equation}
 for $\Gamma_m$. Hence for sufficiently large $M$ the matrix
 of our linear algebraic system contains a 
  non-degenerate Vandermond matrix as a submatrix thus that linear algebraic system is resolvable. 
\vspace{0.2cm}

Using this lemma, we can choose $S_{jm}$ and $M_{jm}$ such that $G_i$ take the form

\begin{equation}\label{preHop6}
     G_i(\tilde X)= \sum_{j=1}^N  \rho K_{ij}  \sigma(\rho^{-1}
     \tilde X_j -h_j) - \lambda \tilde X_i,
\end{equation}
where $\lambda >0$.

\subsection{Control of dynamics for Hopfield system} \label{contrHop}

Using (\ref{preHop6}) we obtain the following system for new variables $Y_i =\rho \tilde X_i$:
\begin{equation}\label{preHop7}
     \frac{dY_i}{dt}= \sum_{j=1}^N  \rho K_{ij}  \sigma(
      Y_j -h_j) - \lambda Y_i.
\end{equation}

It is easy to show that  system (\ref{preHop7})  has a compact attractor.
In fact, $0 \le \sigma(z) \le 1$ thus  that system implies the inequalities 
$$
\frac{dY_i}{dt} \le  N |{\bf K}|   - \lambda Y_i,
$$
where $|{\bf K}|=\max_{i, j} |K_{ij}|$. These differential
inequalities lead to the estimate
$$
|Y_i(t)| \le  (Y_i(0)- N |{\bf K}|\lambda^{-1} ) \exp(  - \lambda t) + N|{\bf K}|\lambda^{-1}.
$$
The last estimate shows that system  (\ref{preHop7}) has an absorbing set ${\mathcal A}=\{ Y: |Y_i| < N |{\bf K}|\lambda^{-1}\}$,
 thus it is dissipative and has a compact attractor.  This result justifies our hypothesis on smallness of kink oscillations at points
 $\bar X_i$ and the transformation of general system (\ref{preHop}) to the Hopfield system (\ref{preHop7}). 
 
 The following claim is proved in \citep{Vak2}.
\\

{\bf Theorem I.}  \label{Comp2}
 { \em  Dynamics defined by system
  (\ref{preHop7})  generates all finite dimensional hyperbolic dynamics (up to orbital topological equivalency) by variations of parameters ${\bf K}, N, \lambda$ and $h$. 
 }

%\end{theorem}

%\subsection{Numerical simulations} \label{numres}

%compareRates

%    \begin{figure*}
% \includegraphics[width=0.5\linewidth]{mutNom2.eps} \label{fig3a}
%{\caption{\small 
%This plot shows cell differentiation as a result of change of a random  %environment impact $\bar \xi(x,y)$, which randomly evolves in time
%$t \in [0, 50]$. 
%That plot corresponds to the time moment $t=20$. 
%}} 
%\end{figure*}

\subsection{Morphogenesis algorithms by waves and physical effects} \label{TuringP}

Let 
 ${\mathcal K}_c=\{1, 2, ..., n_c\}$ be a finite set of cell types. We assume that $j$-th cell occupies a subdomain $\Omega_j =\{[\bar x_i -\delta_c, \bar x_j +\delta_c] \times [0, 1]\} \subset \Omega_L$ 
centered at $\bar x_j$, where $\bar x_j =\delta_c/2 + (j-1)\delta_c$.   We thus have $M=[L/\delta_c]$ equidistant cells, $j=1,...,M$. 
The output cell pattern can be considered as 
 a string $s_{out}$ in the alphabet ${\mathcal K}_c$: $s_{out}=\{a_1 a_2 ... a_M \}$, where $M$ is the number of cells.
 We suppose that condition (\ref{kinkdist}) holds, which means that all kink chain can enter in  the cell. 
 
 Let us describe now how the cell pattern can be produced in our model. We follow classical ideas \citep{Turing, Wolpert}.
 The cell pattern is a result of terminal differentiation, which goes by morphogens.  
Suppose that $u$ is a  morphogen. Following the positional information concept, one can assume that
differentiation starts, when the concentration of $u(\bar x_j, t)$ at the cell center 
is large enough, say, $u(\bar x_j, t)= 1- b$, where $b>0$ is  small. Suppose that
initial coordinates $X_i(0)$ of all the kinks satisfies $X_i(0) < \bar x_1$ and the topological charge
of the leading kink with the coordinate $X_n(t)$ is $1$. Then for the  moment  $t_j$ one has
$t_j \approx (\bar x_j- X_n(0))/\kappa$ (it is a moment when the leading kink approaches the center of $j$-th cell). 
 It is well known that cells interpret morphogen signals by
averaging in space and time. For simplicity, we consider
linear averaging operators
 \begin{equation} \label{TD}
 P_j[u]=\int_{\Omega_j} \omega(x,y) u(x,y, t) dx dy, 
 \end{equation}
  where  $\omega$ is  a smooth  weight function  with the support $\Omega_j$.  
We have 
$u \approx U_N(x, X(t))$. Since for small $\epsilon>0$ the kink chain $U_N$ can be approximated by a piecewise-constant functions with
breaks at $x=X_i$  and kinks oscillate at points $\bar X_i+\kappa t$, relation (\ref{TD}) leads to 
\begin{equation} \label{TD1}
     P_j[u] (t_j) \approx \phi( \tilde X(t_j)) =const + \sum_{j=1}^N   w_l \tilde X_j(t_j),
  \end{equation}
where $w_l$ are coefficients.
Thus we conclude that  the cell can obtain an information about coordinates $X$  at the moment $t_j$, when the waves reach the cell, via
the linear combinations $\phi(\tilde X)$ of the kink positions.

Let us consider how this information can used.
The range of all possible values $X$ will be denoted by
$\Pi$.  We introduce the partition of $\Pi$ consisting of subsets  
$\Pi_k$, $k=0,1, ..., n_c$ with disjoint open interiors such that their union is
$\Pi$. The set 
$\Pi_0$ corresponds to non-differentiated cells, the set 
$\Pi_k$ with $k >0$ corresponds to cells of $k$-th type. 
We encode kink states $X=(X_1, ..., X_N)$ by  functions $Z(X)$. 
The function $Z(X)$ takes the value $k$ if $X \in \Pi_k$.

Then the output string $s_{out}$ of cell types can be  defined as follows:
 $j$-th element of the string is $k$, if $Z(X(t_j))=k$,
 where $t_j$ is the moment when the waves reach $j$-th cell
 (see above).  This construction replaces thresholds in the Wolpert positional information
approach, but in our case this information is transferred in the cells  by waves instead of gradients. 
 Note that at the moment $t=t_j$ the corresponding cell accepts $k$-th state and does not change its type anymore. 
We use here the biological fact that  usually cells 
do not change their cell types after terminal differentiation, when they acquire their specialized type. 

%Let us consider now the case of two cell types, when $n_c=2$ and ${\mathcal K}_c=\{1,2\}$ (the general case can be
%reduced to $n_c=2$ by a binary tree). 

%The key moment is that we can define  coefficients $w_l$ of the averaging functional in such a way that
%$P(X)$ becomes a linear classifier for sets $\Pi_1, \Pi_2$ (like to a single layered perceptron).  It is possible if $\Pi_1$ and $\Pi_2$
%  can be separated by a hyperplane, and one can show (see below) that it is our case. 

%Suppose that at $t=0$ all cells are in an indefinite state $0$. It can achieved if all $X_i(0)$ are sufficiently close
%to $x=0$. 
%It follows from our assumptions to $\delta, \epsilon$ 
%and starting coordinates $X_l(0)$. 

\subsubsection{Algorithm using Markov partitions and Bernoulli shift} \label{Bernovchar}

An algorithm to solve the pattern generation problems is based on 
 celebrated results of  dynamical system theory on hyperbolic sets, its persistence, existence of Markov partitions
 and a connection between Bernoulli shifts and maps on invariant hyperbolic  sets \citep{Katok, Moore1, Moore2}.
 Let us consider a smooth map $q \to G(q)$, where $q$ lies on a smooth compact finite dimensional manifold (for example, torus). Suppose this map defines a dynamical system with discrete time:
 $q(t+ \Delta t)=G(q(t))$, which has  a hyperbolic invariant 
 set $\Gamma$. Dynamics
 on $\Gamma$ can be described by a Markov partition consisting
 of a family of sets $E_j$ \citep{Moore2}.  
 Iterations $q \to G(q)$ is equivalent to a Bernoulli shift map $\Sigma$ defined 
on the set of all such two-sided sequences $(a_j)$:  $\Sigma(a_j)= (a_{j+1})$ \citep{Moore2}. Each points on the hyperbolic set
has an "address", which is a two-sided sequence 
$(a_j)$, $j \in {\mathbb Z}$, and that addresses predetermines 
the point fate under dynamics \citep{Moore2}. Nonetheless such shift dynamics may be chaotic:   a small error in initial data
can lead to an exponential divergence in subsequent iterations, to predict the system $k$
steps in the future, we need to know roughly $k$ symbols  of the initial sequence. 

To use these ideas, 
we first encode cell pattern as a  string in an alphabet ${\mathcal K}_c$ of cell types.
The algorithm input is  then a string $s_{out}=\{a_1 a_2 ... a_M\}$ of symbols from ${\mathcal K}_c$. The algorithm steps are as follows.
 
\begin{enumerate}
 \item We find a discrete time dynamical system
    $q(t+\Delta t) = G(q(t))$ with an appropriate hyperbolic attractor, which has  a Markov partition consisting of $n_c$ subsets
    $E_j$. For the alphabet ${\mathcal K}_c=\{1,2\}$ (or $(red, blue)$) one can
     use the famous map generating a hyperbolic chaotic behaviour, so-called Arnold's cat map defined on the torus
     ${\mathbb T}^2$. Let us define the matrix ${\bf M}$ by
$$
\begin{pmatrix}
2 & 1 \\
1 & 1 
\end{pmatrix}
$$
    and let us set $G(q)={\bf M} q$. The map $q \to G(q)$ is conjugate to a Bernoulli shift
    and the corresponding Markov partition consists of two rectangles
    $E_1, E_2$;
    
    \item By a suspension (see Smale, \citep{smaleBu}) we find an integer $n>0$ and
    a smooth vector
    field $Q(q)$ such that the corresponding flow $S^t$
    defined by the system 
    \begin{equation} \label{maindyneq2} 
    \frac{dq}{dt}=Q(q), 
    \end{equation}
     on a compact domain ${\mathbb B}^n \subset {\mathbb R}^n$ with a smooth boundary
    has a Poincar\'e section  and the corresponding Poincar\'e map  is 
    the map $q \to G(q)$ described at the previous step;

    \item Using equivalency  between iterations $G$ and the Bernoulli shifts we  find a subset $A_0$  consisting
    of initial data  $q(0)$ for (\ref{maindyneq2}) such that the $j$-th  iteration of $G$ enters for the subset
    $E_{a_j}$ for  $j=1, 2, ..., M$;

 \item Let $\epsilon_G >0$ be a constant such that the Poincar\'e map $G$ persists under perturbations of
 the vector field $Q$, which   $\epsilon_G$-small in $C^1$-norm. 
 Such a constant exists due to properties of hyperbolic dynamics (see Appendix  and \citep{Katok, Ruelle1}).  
 We realize the vector field $Q$ by a Hopfield system (\ref{maindyneq}) within accuracy $\epsilon_G$ (about realisations see Appendix); then
 dynamics of $X$-states is defined by the map $q \to X(q)$, where $q$ evolves according to $\epsilon_G$ -perturbed equation
 (\ref{maindyneq2});
 
 \item we define the sets ${\Pi}_k$ in $X$ space  as ranges of rectangles $E_k$ under the maps $q \to X(q)$;
 
 \item We find sufficiently small parameters $\epsilon, \gamma, \kappa>0$ and initial data $z_0$ such that dynamics of kink chain solutions $U_N(x, y, X(t))$ of
 IBVP defined by (\ref{GLE})-(\ref{Hyp}),  (\ref{Neum})-(\ref{initdata}) realizes the Hopfield dynamics found at the previous step;

    \item   we release a kink chain wave   at a suitable speed $\kappa>0$.
    
\end{enumerate}

 The last point of the algorithm is based on the well known biological
 fact that cells (except for stem cells) are not capable of further differentiation.
 When a wave comes to an area occupied by a cell, it changes its type (depending on the amplitude of the wave), that is, it makes differentiation and after that the cell no longer changes.

Note that if $G$ is the Arnold cat map, then one can take
    $n=6$. It follows from Whitney theorem. The strong Whitney embedding theorem states that any smooth $m$-dimensional manifold ( Hausdorff and second-countable) can be smoothly can be embedded in the  $2m$-dimensional Euclidian space.
    The Smale suspension gives us $3$-dimensional manifold, where a flow generates the Arnold map as a Poincar\'e map. 
    This manifold can be embedded in ${\mathbb R}^6$.

%For some values of $\tau$  it is possible 

\subsubsection{Simplified algorithm} \label{simpalg}

Let eq. (\ref{maindyneq2}) define a smooth dynamical system defined on a ball ${\mathbb B}^n$ in ${\mathbb R}^n$  with 
an attractor $\Gamma$, which has an invariant measure $\mu$ defined on $\Gamma$ (for axiom A attractors and Anosov diffeomeorphisms  such measures exist and they are well studied, they are called Sinai–Ruelle–Bowen  (SRB) measures \citep{SRB1,SRB2}).  Suppose that the flow $S^t$ generated by
system  (\ref{maindyneq2}) has the strong mixing property, i.e.
$$
 \mu(S^t A \cap B) \to \mu(A) \mu(B) \quad t \to +\infty 
$$
for two measurable sets $A, B$.  Let $\mu(A_0)$ be a subset of non-zero $\mu$-measure on $\Gamma$ and $V_0$ be a small open neighborhood of $A_0$ in ${\mathbb B}^n$. Let $E_1, E_2,...,   E_{n_c}$ be a fixed partition of $\Gamma$ and $p$ be a fixed positive integer. Then the mixing property implies that if $\Delta T $ is large enough all the following intersections are non-empty: 
$$
  B_{kj}=S^{ j \Delta T} A_0 \cap E_k \ne \emptyset, \quad j=1,..., p, \ k=1,..., n_c, 
$$

In fact, according to the strong mixing property $\mu(B_{kj}) >0$. Moreover, it is easy to see
that 
$$
  V_{kj}=S^{ j \Delta T} V_0 \cap E_k \ne \emptyset, \quad j=1,..., p, \ k=1,..., n_c. 
$$
This shows that for any finite sequence $\{a_j\}$, $j=1,..., p$ of $a_j \in {\mathcal K}_c$ there is 
an open (possibly small) set of initial points $X(0)$ such that
$$
S^{ j \Delta T}(X(0)) \in E_{a_j}, \quad j=1, ..., p.
$$
Let us note 
that the sophisticated algorithm  with Poincar\'e map has an advantage with respect to
the simplified one: by the Bernoulli shifts and the Markov partitions, we can find the set of initial data $q(0)$ in an explicit way.
For simplified algorithm it can be done numerically. We have checked it for the Lorenz system for $q$.

\section{Appendix}  \label{App2}

\subsection{Realisation of vector fields (RVF)} \label{MainR}

The main technical tool in proving attractor complexity for 
partial differential equations and systems is
realization vector field (RVF) method based, in particular, on structural stability ideas. 
It is based on a classical center manifold technique  and on the well known 
idea that any $n$-dimensional dynamics can bifurcate from an equilibrium with $n$-
zero eigenvalues if the number of bifurcation parameters is large enough. Such approach
was used for finite dimensional systems (see \citep{AAIS}), but it can be extended on
infinite dimensional evolution equations. This RVF approach  is developed by
first P. Pol\'a\v cik to prove  existence of non-trivial large time behaviour 
  for quasilinear parabolic  equations (see \citep{Pol2,Pol3}) and developed in \citep{Vak2,Vak16}
  for reaction-diffusion systems, in \citep{switch} for neural networks and in \citep{Vak21} for weakly compressible Navier-Stokes equations.  
  
  In our model, $n=N$, where $N$ is the number of kinks, eigenvalues are exponentially close to zero 
  and the initial data $z_0$ plays the role of the main bifurcation parameter, i.e., the bifurcation parameter is  infinite dimensional. 

\subsubsection{Structural stability}

Recall the basic concept of structural stability introduced by A. Andronov and S. Pontryagin in 1937.  Consider  a smooth vector field $Q$ on  compact domain ${\mathbb D}^n$ of $\mathbb{R}^n$ with a smooth boundary  (or on a compact smooth manifold $M$ of dimension $n$). Assume that $Q \in C^1({\mathbb D}^n)$ and consider all $\delta$-small perturbations
 $\tilde Q$ such that
\begin{equation}\label{strucstab}
 |\tilde Q|_{C^1({\mathbb D}^n)} <  \delta.
\end{equation}

Consider systems of differential equations $dq/dt=Q(q)$ and $dq/dt=Q(q) + \tilde Q(q)$ and suppose that they define global semiflows $S_Q^t$ and $S_{Q +\tilde Q}^t$ on ${\mathbb D}^n$. The system $dq/dt=Q(q)$ is called structurally stable if there exists a $\delta_0$ such that if
$$
|\tilde Q|_{C^1( {\mathbb D}^n )} < \delta_0, 
$$
then
trajectories of semiflows $S_Q^t$ and $S_{Q +\tilde Q}^t$ are orbitally topologically conjugate (there exists a homeomorphism, which maps trajectories of the first system into trajectories of the second one). Roughly speaking, the original system is structurally stable if any sufficiently small $C^1$ perturbations of that system conserve the topological structure of its trajectories, for example, the equilibrium point stays an equilibrium (maybe, slightly shifted with respect to the equilibrium of non-perturbed system),  the perturbed cycle is again a cycle (maybe, slightly deformed and shifted).  

Note that structurally stable dynamics may be, in a sense, "chaotic". There is a rather wide variation in different definitions of "chaos". We restrict ourselves hyperbolic chaotic sets.  Chaotic
(no periodic and no rest point) hyperbolic sets occur in some model systems 
\citep{Smale,Ruelle1, Katok}.

\subsection{RVF for evolution problems in Banach spaces}
\label{sec:3}

Let us consider a family of local semiflows $S^t_{\mathcal P}$ in a fixed Banach
space $B$. Assume
 these semiflows  depend on a parameter ${\mathcal P} \in B_1$, where
$B_1$  is another Banach space.
Denote by ${\mathcal B}^n(R)$ the  ball $\{q: |q|\le R\}$ in ${\mathbb R}^n$, where $q=(q_1, q_2, ..., q_n)$ and
$ |q|^2=q_1^2 +
... + q^2_n$.  For $R=1$ we will omit the radius $R$,  ${\mathcal B}^n={\mathcal B}^n(1)$. Remind that 
a  set $M$ is said to be locally invariant in an open set $W \subset B$  under a semiflow $S^t$ in $B$ if  $M$ is a subset of $W$ and each trajectories of $S^t$ leaving $M$ simultaneously leaves
$W$.  In this paper, all $W$ are tubular neighborhoods of the balls ${\mathcal B}^n(R)$, which have small widths.
Consider system of differential equations defined on the ball ${\mathcal B}^n$:
\begin{equation}
 \frac{dq}{dt}=Q(q),
\label{ordeq}
\end{equation}
 where
\begin{equation}
   Q \in C^1({\mathcal B}^n), \quad \sup_{q \in {\mathcal B}^n}|\nabla Q(q)| < 1.
\label{cond1}
\end{equation}
 Assume the vector field $Q$ is directed strictly
inward at the boundary $\partial {\mathcal B}^n=\{q: |q|=1 \}$:
\begin{equation}
   Q(q)\cdot q < 0 , \quad  q \in \partial {\mathcal B}^n.
\label{inward}
\end{equation}
Then system (\ref{ordeq}) defines a
global semiflow
 on ${\mathcal B}^n$. Let $\delta$ be a positive number.

{\bf Definition.} ({\bf realization of vector fields})  \label{RealVF} 
{\em We say that the family of local  semiflows $S^t_{\mathcal P}$  realizes the vector field $Q$ (dynamics (\ref{ordeq})) with accuracy $\delta>0$
(briefly, $\delta$  - realizes),
if there exists a  parameter ${\mathcal P}={\mathcal P}(Q, \delta, n)  \in B_1$
such that

({\bf i}) semiflow $S^t_{\mathcal P}$ has a locally invariant   in a open domain ${\mathcal W} \subset B$ and locally attracting
 manifold ${\mathcal M}_n \subset B$ diffeomorphic to the unit ball $ {\mathcal B}^n$;

({\bf ii})  this manifold is embedded into $B$  by a  map
\begin{equation}
  z = Z(q), \quad q \in {\mathcal B}^n, \quad z \in B, \quad Z \in C^{1+r}
({\mathcal B}^n),
\label{manifRVF}
\end{equation}
where $r > 0$;

({\bf iii}) the restriction of the semiflow  $S^t_{\mathcal P}$ to ${\mathcal M}_n$  is defined by the system of differential equations
\begin{equation}
  \frac{dq}{dt}=Q(q) + \tilde Q(q, {\mathcal P}), \quad Q \in C^{1}({\mathcal B}^n),
\label{reddynam}
\end{equation}
  where
\begin{equation}
	    |\tilde Q(\cdot, {\mathcal P})|_{C^1({\mathcal B}^n)} <
\delta.
\label{estRVF}
\end{equation}}
%\end{definition}

%This means that the dynamics on the invariant manifold
%is defined by   the variables $q_1, q_2, ..., q_n$ and
%approximates prescribed dynamics (\ref{ordeq}) with
%an accuracy $\epsilon$.

{\bf Definition.}  \label{RVFmax} 
{\em Let $\Phi$ be a set of vector fields $Q$, where each $Q$ is defined on a ball ${\mathcal B}^n$, positive integers $n$ may be different.  
We say that the family ${\mathcal F}$ of local semiflows $S^t_{\mathcal P} $ realizes the family $\Phi$ if for each $\delta>0$ and each $Q  \in \Phi$ the filed $Q$ can be 
$\delta$
-realized by the family ${\mathcal F}$.

We say that the family of global semiflows ${\mathcal F}$ has the property of universal dynamical approximation if that family realizes the set of all $C^1$- smooth finite dimensional fields defined on all
unit balls ${\mathcal B}^n$. 
}

Many systems enjoy the property of universal dynamical approximation, for example,  the Lotka-Volterra system with many species, 
the Hopfield system, a large class of reaction-diffusion systems and others. 

\subsection{Estimate of accuracy of asymptotic wave solutions} \label{linop}

In this subsection, we estimate the accuracy of asymptotic
solutions, which are correct under some restrictions to parameters  $\epsilon, \gamma$ and $\kappa$.
We consider the IBVP defined by  \eqref{OB1W}-\eqref{OB2W}, boundary  conditions (\ref{Dir}), (\ref{period}), 
(\ref{Neum}) and initial data (\ref{initdata}) assuming that initial data for $u$ lie in a narrow neighborhood
of the kink chain $U_N$.  Then, by standard semigroup theory \citep{Henry},  we are capable to prove the global existence of solutions 
for all $t>0$ and justify correctness of kink chain asymptotics. The statement mainly follows 
\citep{Carr, Vak2, Vak16}.

\subsubsection{Linear operator and projections} \label{Linproj}

Let us consider the linear operator ${\bf L}_{\epsilon}$
associated with the linear part of eq. (\ref{GLE}) for $u$ and defined by (\ref{operator}). 
This operator depends on kink coordinates $X$ as a parameter, however, under condition (\ref{kinkdist}) we can  obtain estimates of its spectrum uniform in $X$. Taking into account this fact, we omit  a dependence on $X$ in notation. 

Let us introduce the complementary projection operators
defined on $L_2(\Omega)$:
$$
{\bf P}_{\epsilon} u=\sum_{j=1}^N \langle u, \psi_j \rangle 
\psi_j, \quad {\bf Q}_{\epsilon}={\bf I} -{\bf P}_{\epsilon}. 
$$

Let us formulated an auxiliary lemma. 

{\bf Spectral Barrier Lemma.}
{\em Let $X$ satisfy (\ref{kinkdist}). Let $u \in H(\Omega)=W_{2,2}(\Omega)$ and 
$
{\bf P}_{\epsilon} u=0$.
Then for sufficiently small $\epsilon>0$ 
\begin{equation} \label{Spectr}
 || {\bf L}_{\epsilon} u|| \le - c_0 \epsilon^{2}|| u||. \ 
\end{equation} 
}
This assertion is simply a reformulation of Lemma I by projection operators. 

So, if we restrict the operator ${\bf L}_{\epsilon}$
 to functions orthogonal to all $\Psi_j$, then for that operator there exists a small spectral barrier of the order 
$\epsilon^2$.  
In subsequent estimates this fact plays a key role.
Moreover, constants $c_1, c_2, C_1, ...$, which appear
in those estimates, are uniform in small parameters
$\gamma, \epsilon$ and $\kappa$.

\subsubsection{Function spaces, norms and estimates} \label{normspace}

  Let us introduce  the inner scalar product in the space of $2\pi$ -periodic in $x$ measurable functions defined on
$\Omega$ by 
\begin{equation}
  \langle u,  w \rangle =\int_0^{2\pi} \int_0^1  u(x,y) w(x,y) dxdy.
\label{norm}
\end{equation}
Let  $||u||$ be the corresponding norm, i.e., $||u||^2=\langle u, u\rangle$. We denote by $H$ the Hilbert space  of measurable
functions defined on $\Omega$ and $2\pi$- periodical in $x$
 with bounded  norms $|| \ ||$.
We consider our IBVP problem  in the space   ${\bf H}=H \times H$, i.e., $u \in H$ and ${v} \in H$.

Let us introduce the  fractional spaces \citep{Henry} defined by
$$
H_{\alpha}=\{   v \in H: \quad ||(-\Delta_D)^{\alpha}v ||=||v||_{\alpha} < \infty \    \},
 $$
where $\Delta_D$ is the Laplace operator under the Dirichlet boundary conditions with  a natural definition domain
$
Dom \Delta_D$
and $\alpha \ge 0$. Here $H_{0}=H$.
Similarly,
$$
\tilde H_{\alpha}=\{u \in H: \quad ||(-\Delta_N)^{\alpha} u ||=||u||_{\alpha} < \infty \},
 $$
where $\Delta_N$ is the Laplace operator under the Neumann boundary conditions. 
 We  denote the product $\tilde H_{\alpha} \times  H_{\alpha}^m$ by ${\bf H}_{\alpha}$.

   Let us introduce the corresponding fractional spaces
$$
H_{\alpha}= \{ { u} \in H:  \quad ||({\bf I}-\Delta)^{\alpha} { u}|| < \infty \}
$$
with the norms
$$
|| u||_{\alpha}=||(I - \Delta)^{\alpha} u||,
$$
where $\alpha \in (0, 1)$.
We  use the well known estimate \citep{Henry}
\begin{equation} \label{eq3int}
||{  u}^2|| \le c_1 || { u} ||_{\alpha} ||{ u}||,
\end{equation}
where $c_1(\gamma) >0$ is a constant.

Let us prove first that our IBVP problem is well posed and defines a local semiflow. The proof is standard  and follows
\citep{Henry}. We consider this problem in the Hilbert phase space ${\bf H}$. Let  ${\bf v}=({u}, v)^{tr}$ and $||{\bf v}||=||u|| + ||{v}||$.
Our IBVP  can be represented as an evolution equation  \citep{Henry}
\begin{equation}
  {\bf v}_t = A{\bf v} + F({\bf v}),
\label{eqfluid}
\end{equation}
  where  $A=(\epsilon^2 \Delta_N,  D \Delta_D)^{tr}$ is a 
self-adjoint operator in ${\bf H}$ and
$$
  F=\big(u - u^3 + \gamma v, \   z_0 u + \kappa v_x \big)^{tr}.
$$

We use the Sobolev embeddings
\begin{equation} \label{embed}
  || u||_{L^{\infty}(\Omega)} \le c_0 || u||_{\alpha},
  \end{equation}
that gives
$$
  ||{\bf v} ||_{L^{\infty}} \le  c_2 || {\bf v}||_{{\bf H}_{\alpha}}=||{\bf v}||_{\alpha}.
$$
These estimates show that $F$ is a   $C^{1}$-map from ${\bf H}^{\alpha}$ to ${\bf H}$ and thus  
 eq. (\ref{eqfluid}) defines a local semiflow \citep{Henry}. 
%\vspace{0.2cm}

\subsubsection{Global existence} \label{globex}

To establish existence of bounded solutions of our IBVP 
on infinite time interval $(0, \infty)$, we need a priori estimates of weak norms, for example, $||u||(t)$ and $||v||(t)$. 
They can be obtained in a standard way.  Let us consider  scalar products of the left and right hand sides of 
\eqref{GLE} with $u$. 
Then under our boundary conditions eq. (\ref{GLE}) implies the estimate
\begin{equation} \label{uest}
    \frac{d||u||^2}{2dt} \le - \frac{\epsilon^2}{2} || \nabla u||^2 + C_0 +  \gamma ||v||||u||.
\end{equation}
The same trick for eq.(\ref{OB2M}) gives
$$
\frac{d||v||^2}{2dt} \le - || \nabla v ||^2 + \langle z u_x,v \rangle.
$$
By integrating by parts in the last term  and using that $z$ is a smooth and bounded  function
one obtains
$$
\frac{d||v||^2}{2dt} \le -  || \nabla v ||^2 + c_1 || u|| ||v||.
$$
Then the last differential inequality and the Poincar\'e inequality give
$$
\frac{d||v||^2}{2dt} \le -  c_0 ||  v ||^2 + c_1 (\sup_{s \in[0,t]} ||u||(s)) ||v||, \ t\in I_T,
$$
where $I_T=[0, T]$.
Thus we have the estimate $||v||$:
\begin{equation} \label{vest}
  ||v||(t) \le c_1 \big(||v||(0) + \sup_{s \in[0,t]} ||u||(s) \big),
  \ t\in I_T.  
\end{equation}
We substitute this estimate into (\ref{uest})   that leads to
\begin{equation} \label{uest2}
    \frac{d||u||^2}{2dt} \le  C_0 + c_4 \gamma  (C_1^2 +  \sup_{s \in[0,t]} ||u||(s))||u||).
\end{equation}
for $t\in I_T$. Consider the differential equation
$$
\frac{dZ^2}{2dt} =C_0 + 1 + c_4 \gamma  (C_1^2 +  Z^2).
$$
We observe that if $Z(0) > ||u(0)||$ then $Z(t) > ||u(t)||$ for all $t >0$. The differential equation
for $z$ can be rewritten as
$$
\frac{dZ}{dt} =(C_0 +1 + c_4\gamma  C_1^2) Z^{-1}+ 1 + c_4 Z.
$$
We observe then that $Z \le \max{1, Y}$, where $Y$ is a solution of the linear differential equation
$$
\frac{dY}{dt} =(C_0 + c_4\gamma  C_1^2)  + 1 + c_4 \gamma Y.
$$
Therefore, we conclude that
$$
Z(t) < \max\{1,  C_5 \exp(\gamma t)  \}
$$
and thus
\begin{equation} \label{uest2b}
    ||u||^2(t) \le   \max\{1,  C_5 \exp(\gamma t)  \}.
\end{equation}
By (\ref{vest}) this estimate implies 
\begin{equation} \label{uest2b2}
    || v||^2(t) \le  c_2\big( \max\{1,  C_5 \exp(\gamma t)  \} + ||v||(0) \big).
\end{equation}

These estimates show that the norms of solutions are bounded
on all bounded time intervals $I_T$ although these norms may slowly increase in $t$.
It implies, together with the estimates of the previous subsection, that solutions of our IBVP exist for all positive
times and unique. Therefore, our IBVP generates the global semiflow.  

In the coming subsections we show, in particular, that
if the initial data for $u$ are close to the 
kink chain  then the norms $||u||$ and $||v||$ stay bounded for all times while the mutual kink distances
stay more than a small $\delta_1>0$ (for sufficiently small positive $\epsilon << 
\epsilon_0(\delta_1)$.  If the kink coordinates
oscillate at $\bar X$ remaining in a small $\rho$-neihborhood
then $||u||$ and $||v||$ are bounded for all times. So,
if the kink dynamics is governed by the Hopfield system
then the norms $||u||$ and $||v||$ are bounded for all times.

\subsubsection{Representation of solutions}  \label{repr}

We represent our solutions as
\begin{equation} \label{repau}
u=U_N(X) + \tilde u,
\end{equation}
\begin{equation} \label{repav}
v=V_N(X) + \tilde v,
\end{equation}
where $V_N$ are defined by  (\ref{VN}), (\ref{Wj}) and 
$$
{\bf Q}_{\epsilon}\tilde u =0, 
$$
In coming subsection we derive dynamical equations for new unknown variables for
$X$, $\tilde u$ and $\tilde v$.

\subsubsection{Equations for $X$, $\tilde u$ and $\tilde v$} \label{eqtueqtv}

The estimates of this subsection hold under condition

We substitute (\ref{repau}) and (\ref{repav}) into
(\ref{GLE}) and (\ref{OB2M}) that gives the system 
\begin{equation} \label{utilde}
    \tilde u_t +\bar  c_j^{-1} (-1)^i \epsilon^{-1/2} \sum_{j=1}^N \frac{dX_j}{dt} \Psi_j={\bf L}_{\epsilon}\tilde u + F_j(X, \tilde u,  \tilde v),
    \end{equation}
   \begin{equation} \label{vtilde}
    \tilde v_t =D\Delta \tilde v  + z_0 \tilde u + \kappa \tilde v_x -\sum_{i=1}^N \frac{\partial V_N}{\partial X_i}\frac{dX_i}{dt},
\end{equation} 
     where
  \begin{equation} \label{Fj}  
    F_j= - 3 U_N \tilde u^2 -\tilde u^3 + \gamma (V_N(X) +\tilde v).  
    \end{equation}

To find equations for the slow variables $X_i$ and the fast 
ones $\tilde u, \tilde v$ we apply to (\ref{utilde}) our projection operators that allows us to represent the system in the standard
slow-fast form. This procedure gives 
\begin{equation} \label{utildeP}
    \tilde u_t =
    {\bf L}_{\epsilon} \tilde u  + {\bf Q}_{\epsilon}F_j(X, \tilde u, \tilde v)   
    \end{equation}
    \begin{equation} \label{utildeX}
    \frac{dX_i}{dt}=G_i(X, \tilde u, \tilde v),
    \end{equation}
    where 
$$
G_i(X,  \tilde u, \tilde v) = (-1)^i \epsilon^{1/2} \bar c_i \langle F_i, \Psi_i \rangle.
$$
    Using these equations   we are capable  to estimate
$\tilde u$ and $\tilde v$ by the standard semigroup theory.

\subsubsection{Estimates of $\tilde u$ and $\tilde v$} \label{tutv}

Let $t \in [0, T]$, where $T$ may be large. All 
the subsequent estimates are uniform in $T$ under  restrictions (\ref{asseps}) to our small parameters.
We introduce the norms
$$
||| \tilde u |||_{\alpha} =\sup_{t \in [0, T]} ||| \tilde u|||_{\alpha},
\quad  ||| \tilde v |||_{\alpha} =\sup_{t \in [0, T]} |||\tilde  v|||_{\alpha},
$$
and for $\alpha=0$ we write down simply $|||u|||$ etc.

Using the Spectral Barrier Lemma,  we obtain, in a standard way by semigroup estimates \citep{Henry, Vak16}, that
$$
|||\tilde u|||_{\alpha} \le 
C_1 \epsilon^{-2} |||F_j||| + C_M ||\tilde u(0)||_{\alpha}, 
$$
where $C_1, C_M >0$  are constants uniform in $\epsilon$.  We note that  
$$
|||F_j|||\le C_2 \Big(\gamma \big ( |||V_N  ||| + |||\tilde v||| \big) +  \sup |\tilde u| |||\tilde u|||\Big),
$$
where the supremum of $|u(x,y, t)$ is taken over all $x,y \in \Omega$ and $t \in [0, T]$.
We observe that $|||V_N||| < C_3$.  We also use the estimate (\ref{embed})  for $\alpha > 3/4$ that gives
\begin{equation} \label{estuF}
|||u|||_{\alpha}\le C_3 \epsilon^{-2} \Big(\gamma \big ( C_3  + |||\tilde v||| \big) +   |||\tilde u|||_{\alpha}^2 + ||\tilde v||(0)\Big).
\end{equation}

Similarly,   
\begin{equation} \label{estvF}
|||\tilde v|||_{\alpha}\le C_4  \big(|||\tilde u|||_{\alpha} + |dX/dt| + ||\tilde v||(0)\big).
\end{equation}
 The term $|dX/dt|$ can be estimated by (\ref{utildeX}). We find that
 \begin{equation} \label{estXF}
|dX/dt| < C_5 \epsilon^{1/2} \big(|||\tilde u|||_{\alpha}^2 + |||\tilde u|||_{\alpha}^3 +\gamma (C_6 + |||\tilde v|||_{\alpha})\big).
\end{equation}
 The system of inequalities (\ref{estuF}), (\ref{estvF}), (\ref{estXF})   implies that under condition $\gamma << \epsilon^4$ our evolution problem is weakly nonlinear and for small 
$||\tilde u||(0), ||\tilde v||(0) << c\gamma$ it can be resolved by standard contracting map principle \citep{Henry}.
We obtain
\begin{equation} \label{tildeuest}
|||\tilde u|||_{\alpha} \le 
C_5 \gamma \epsilon^{-2},
\end{equation}
and
\begin{equation} \label{tildevest}
|||\tilde v|||_{\alpha} \le C_5  \gamma \epsilon^{-1/2}.
\end{equation}

\subsubsection{Locally invariant and locally attracting
manifolds}

Let us introduce the domains 
\begin{equation} \label{estXX} 
    {\mathcal D}_R=\{X \in {\mathbb R}^N: \ |X| < R \}.
\end{equation}
Estimates established in the previous section hold under conditions 
\begin{equation} \label{estXX2} 
 X \in {\mathcal D}_R,  \quad dist(X) > \delta_0, 
\end{equation}
where $R$ is a  positive constant.  In general, it is impossible to guarantee that
solutions of system of differential equations  (\ref{utildeX})
satisfy
an  uniform estimate  (\ref{estXX})  for all $t$.
To overcome this difficulty, we perform the well known truncation procedure.  Let $\chi_{R}(x)$ be 
$C^{\infty}$ smooth increasing function of $x \in {\mathbb R}$ such that $\chi_{R}(x) = 1$ for
$|x |<  R$  and  $\chi_{R}(x)=0$ for $|x| > 2R$.  Consider equations 
\begin{equation}
   \frac{dX_i}{dt}  =  \chi_{R_0}(|X|) G_i(X,, \tilde u)= \tilde G_i(X,   \tilde{{v}} , \tilde u).
\label{OB3bT}
\end{equation} 
If $ |X(0)| < 2R_0$ solutions of the Cauchy problem for this system is defined for all $t \in (-\infty, +\infty)$ for any $\tilde{{v}}(t)$ and  $w(t)$.

\vspace{0.2cm}
{\bf Lemma LIM} (on existence of a locally invariant manifold) \label{Lem4.4}
{\em  Let $R_0 >0$ be an arbitrary positive number. Then for sufficiently small positive $\epsilon$
the semiflow, generated by (\ref{utildeP}), (\ref{vtilde}) and (\ref{OB3bT}), has an invariant and locally attracting normally
hyperbolic manifold ${\mathcal M}_N$ of dimension $N$ defined by
\begin{equation}
   {\tilde v}=\gamma { \hat V} (X, \epsilon) 
\label{manv}
\end{equation}
\begin{equation}
   \tilde u= { } \gamma \hat U(X, \epsilon), 
\label{manu}
\end{equation}
where 
 respectively, $\hat {{V}},  \hat U$  are $C^{1+r}$ maps from the ball $\{X: |X| < R_0 \}$ , the number $r \in (0,1)$ and 
 the maps $\hat {{V}},  \hat U$ are bounded in $C^1 $-norm:
\begin{equation} \label{WW}
||\hat U||_{\alpha} \le C_1\gamma \epsilon^{-2}, \quad |||D_X \hat U|||_{\alpha} \le C_2\gamma \epsilon^{-2}, 
\end{equation}
\begin{equation}
||  \hat {{V}} ||_{\alpha} \le C_3\gamma \epsilon^{-1/2}, \quad |||D_X  \hat {{V}}|||_{\alpha} \le C_4\gamma \epsilon^{-1/2}, \label{VV}
\end{equation}
where $C_i>0$ are constants.
} 
\vspace{0.2cm}

{\bf Proof}.
The proof uses  the standard technique, see Theorem 6.1.7  from \citep{Henry} and follows \citep{Vak16}. 
Now we are ready to describe an algorithm to construct
waves with a prescribed large time behavior.

\subsection{Algorithm to find waves having a prescribed 
attractor} \label{alg}

Suppose we would like to have waves having a prescribed
structurally stable  \footnote{If we approximate a dynamics within a finite time interval, the assumption on structural stability
can be removed} attractor   defined by the system (\ref{ordeq}),where the vector field $Q$ satisfies (\ref{cond1}) and
(\ref{inward}).

The algorithm proceeds in two steps, at each step we construct
realizations by fast-slow systems using the RVF method.

{\bf Step 1}.

Using results \citep{Vak2} we $\epsilon_1$-realize system (\ref{ordeq}) by an Hopfield system
of a larger dimension $N > n$.

{\bf Step 2}. We  adjust parameters 
$\epsilon, \gamma, \kappa$ and the radius $R_0$
 to provide
existence of locally invariant and locally attracting manifold
from Lemma LIM.  

Then our IBVP has a local attractor topologically equivalent
to the prescribed one. It can explained as follows.

\subsubsection{Correctness of procedure} \label{corr}

The correctness of that  procedure can be 
demonstrated as follows.

The prescribed structurally stable attractor  is  a compact set in the ball ${\mathbb B}^n$. 
Due to structural stability, 
for sufficiently small $\epsilon_1>0$ this fact implies that a $X$-system for kink motion, which $\epsilon_1$-realizes system
(\ref{ordeq}), also has an attractor contained in a 
ball of a radius $R_0$. Therefore,  $X$-trajectories lie
in that ball and we use that $R_0$  in Lemma on locally invariant
and locally attracting manifold. That manifold has an open
in our phase space attraction basin ${\mathcal B}$.
As it was mentioned above, 
in general the semiflow $S^t$ defined by our IBVP is local in time. However, 
for initial data ${\bf v}(0)$ lying in the attraction basin ${\mathcal B}$, the corresponding 
trajectories ${\bf v}(t)$ approach the locally invariant
manifold and they do not leave a  small neighborhood ${\mathcal W}$ of that manifold while $|X(t)| < R_0$. But 
if $\epsilon_1, \epsilon$ are small enough and all parameters are chosen as above,  the bound  $|X(t)| < R_0$ 
 holds for all $t$. In fact, the $X$ trajectories are defined
 by the Hopfield system, which is dissipative. Therefore these trajectories are bounded. Moreover, due to our specific choice of the Hopfield system we can use  condition (\ref{Osc}), where $|\tilde X_i=O(\rho)$ and $\rho>0$ is small. Hence the kinks do not approach each other
 and we can use our kink chain solution for all times.

%\bibliography{referenceT}

%\bibliography{planetary}

\end{document}